# STRUCTURAL-DAMAGE DETECTION BY DISTRIBUTED PIEZOELECTRIC TRANSDUCERS AND TUNED ELECTRIC CIRCUITS

F. dell'Isola, F. Vestroni, S. Vidoli

Università di Roma ''La Sapienza,'' Dip. Ingegneria Strutturale e Geotecnica, Rome, Italy

*A novel technique for damage detection of structures is introduced and discussed. It is based on purely electric measurements of the state variables of an electric network coupled to the main structure through a distributed set of piezoelectric patches. The constitutive parameters of this auxiliary network are optimized to increase the sensitivity of global measurements— as the frequency, response functions relative to selected electric degrees of freedom—with respect to a given class of variations in the structural–mechanical properties. Because the proposed method is based on purely electric input and output measurements, it allows for accurate results in the identification and localization of damages. Use of the electric frequency-response function to identify the mechanical damage leads to nonconvex optimization problems; therefore the proposed sensitivity-enhanced identification procedure becomes computationally efficient if an a priori knowledge about the damage is available.*

## INTRODUCTION

The problem of structural-health monitoring is one of the most urgent engineering tasks associated with the high requirements and the severe operating conditions imposed on contemporary advanced structures. This subject has received considerable attention in recent years in the literature on aerospace, civil, and mechanical engineering. There is no general, universal approach that could be used to effectively solve this problem in any case; the efficiency of the method depends essentially on the specific structure under consideration, the availability of suitable experimental tests, and the type of occurring damage. Many approaches are based on a direct inspection of the structure in the vicinity of damage; to this class belong methods based on acoustic and ultrasonic measurements, thermal emissions, radiography, and others. Many of them require in addition an a priori knowledge of the domain where the damage occurred and can be used to detect damages on or close to the structure surface. An additional requirement that often eliminates effective approaches from practical engineering applications is

Address correspondence to F. dell'Isola, Università di Roma ''La Sapienza,'' Dip. Ingegneria Strutturale e Geotecnica, Via Eudossiana 18, 00184 Rome, Italy. E-mail: francesco.dellisola@uniroma1.it

the exact and complete knowledge of the state fields, as displacements, stresses, temperatures, in the whole domain of a damaged structure [1].

To detect the occurrence of damaged zones, many approaches have been adopted for the choices of both the forcing inputs and the analysis of the system response. Some authors (see Refs. 2–4 and the references therein) used the information collected in the frequency-response functions; others, see for instance Ref. 5, prefer the use of the time signals; a last tendency is to use wavelet transforms to analyze both the time and frequency contents of the signal [6]. The application of wave-propagation analysis for detecting structural damages dates back to the late eighties and early nineties when the use of stationary waves and of frequency-response functions was already a standard practice. The literature on the subject is extremely wide; a review of the system-identification methods can be found in Ref. 1.

Thus, there is still the need for nondestructive methods enabling damage detection and identification in complex structures; a favorable method should be based on simple, practicable measurements. Among other possibilities, the measurements of natural frequencies (see for instance Refs. 2–4) and local measurements of displacement, strain, or stress values in selected sites of a structure belong to the simplest solutions worthy of interest. Unfortunately such a choice, in spite of significant advantages, is also associated with important difficulties. As a matter of fact damage often results in local mechanical changes of the structural parameters and its early detection is necessary before the severity of damage could lead to abnormal functioning and structural failures. On the other hand these local changes, especially in the early stage, often have little influence on the mechanical global characteristics, as the natural vibration frequencies. The detailed knowledge of displacements or other fields carries information valuable in damage identification and can notably improve the situation but; as it was already mentioned, its actual achievement is either not possible or too expensive (e.g., measurements of displacement fields in complex structures). Even in the case of few lumped external forces, when the determination of the external work is easily achieved through displacement measurements is selected points of the structure, both the eigenfrequencies and the extrernal work represent global quantities and the information concerning the local changes of structural characteristics is deeply ''hidden'' and not directly available. This is why usually global measurements are less sensitive to local moderate variations—such the ones resulting from damages—of structural characteristics.

To avoid the discussed difficulties, the basic problem to be solved is how to increase the sensitivity of selected global measurements up to a level enabling effective damage detection and identification; this problem is addressed with different perspectives in Refs. 7–10. In particular, the ideas proposed in Ref. 9 follow the observation that, by a proper modification of energy distribution in a structure, one can significantly magnify the effects of damage; in the performed theoretical and experimental investigations on simple structural elements as beams and plates, the effect of local structural stiffness variations is enlarged by a factor of 10. This task can be achieved by coupling the main structure to auxiliary systems whose parameters are tuned according to suitable optimal conditions; to this end flexible supports, concentrated masses, discrete vibration absorbers with variable position and mechanical characteristics have been used and tested.

In this article we propose to substitute these mechanical auxiliary systems with purely electric devices: the coupling between the mechanical (main structure) and the electric (auxiliary electric system) subsystems is assured by piezoelectric patches glued along the structure. Indeed in the past few years some novel electromechanical integrated systems have been introduced (see, e.g., Ref. 11) based on the concept of electric analog of a given structure and aimed to control its mechanical vibrations. These structures can be labeled as piezoelectro-mechanical (PEM) because the control of mechanical vibrations is achieved through an electric net connecting a set of distributed piezoelectric patches. Thus, a PEM structure is constituted by a structural member to be controlled, a set of actuators uniformly distributed on the considered structural member, and a suitable electric circuit including as elements the piezoelectric transducers and completed by optimally inserted impedances. Preliminary encouraging results are reported in Ref. 12 where a comparison between the presently proposed technique and a standard approach, measuring the structural eigenfrequencies, is also provided.

Because in the process of increasing the sensitivity we are led to use the electric frequency-response function, the identification procedure involves the simultaneous global minimization (with respect to mechanical-damage parameters) and global maximization (with respect to the electric sensitivity-enhancing parameters of a nonconvex function. If an a priori coarse knowledge about the damage—especially on its localization—is available that sufficiently restricts the range of damage parameters, then the proposed sensitivity-enhanced procedure is computationally efficient; otherwise, the possible existence of multiple optimal points could require heavy computational efforts.

## SENSITIVITY AND EFFECTIVENESS OF DAMAGE-EVALUATION FUNCTIONS

Our approach is actually based on a parametric identification; let us briefly summarize the analogies between the system-identification and the damage-detection techniques. Usually in system identification methods, one measures the response $O_*$ of a given system (i.e., with given actual values $\pi_*$ of parameters) to a forcing input $I_*$:

$$I_* \to \boxed{\pi_*} \to O_*, \quad I_* \to \boxed{\pi} \to O \tag{1}$$

Then, being able to compute the response $O$ of the same system with a generic value $\pi$ of its parameters, one seeks for the value $\pi_* \in \Pi$ that fits $O$ to $O_*$ better than others. Here $\Pi$ means the space of admissible parameters values. Thus, it is rather natural to formulate the system-identification problems as minimization problems of suitable functions: indeed, consider the functional $\mathfrak{E}(\pi; I_*, O_*)$, mapping an admissible value of system parameters $\pi \in \Pi$, and the actual inputs $I_*$ and system responses $O_*$, into real positive numbers such that

$$\mathfrak{E}(\pi_*; I_*, O_*) = 0, \quad \mathfrak{E}(\pi; I_*, O_*) > 0, \text{ for } \pi \neq \pi_* \tag{2}$$

Clearly the identification of the actual value $\pi_*$ is equivalent to find the global minimum of $\mathfrak{E}(\cdot; I_*, O_*)$ in $\Pi$.

In a very similar way, in damage-detection problems one measures the responses $O_0$ and $O_*$ of the same system in two different instants characterized by possibly different values $\pi_0$ and $\pi_*$ of the parameters.; the subscript 0 and * respectively mean the undamaged and actual, possibly damaged, situation. The problem is to identify the variation $\Delta \pi_* := \pi_* - \pi_0$ of the system parameters to which the variation of the system response $O_* - O_0$ is amenable. To this end one must be able to compute the variation $O - O_0$ of the system response associated to a generic variation $\Delta \pi := \pi - \pi_0$ of the system parameters.

$$\begin{cases} I_* \longrightarrow \boxed{\pi_0} \longrightarrow O_0, \\ I_* \longrightarrow \boxed{\pi_0 + \Delta \pi_*} \longrightarrow O_*, \end{cases} \quad I_* \longrightarrow \boxed{\pi_0 + \Delta \pi} \longrightarrow O \tag{3}$$

Hence also the damage-detection problems can be reformulated as minimization problems of suitable functions, the choice of a distinguished function being the selective criterion among different methods. Note that in this process the identification of the actual values $(\pi_0, \pi_0 + \Delta \pi_*)$ of the system parameters is not strictly necessary because only the variations from a fixed reference state, namely the undamaged one, can be compared.

The measured responses $(O_0, O_*)$ can be, in general, modal quantities or time histories of the system-state variables in selected sites of the structure; more often, measurements of the frequency response functions of these sites are used. Thus, for instance, the well-established technique of damage identification through eigenfrequencies measurements can be regarded as the minimization problem for the function summing the squares of the differences between the positions of the peaks of the frequency-response functions $O_0$ and $O_0 + \Delta O_*$ while disregarding any other information about them. Much better results can be achieved considering functions that weigh more information contained in the frequency-response functions, for instance, their values in several fixed frequencies. As remarked in the Introduction, this can be a difficult or expensive task when dealing with mechanical measurements, particularly in complex structures. However, let us explicitly remark that the same task is easily achieved when dealing with measurements on electric systems.

## Possible Choices for the Function 𝔈

In the system identification, a crucial role is played by the choice of the function $\mathfrak{E}(\pi; I_*, O_*)$ to be minimized over the space $\Pi$ of admissible parameters. Here the procedure for parametric identification introduced in Ref. 13 is applied. To this aim, let the equation

$$D(\omega, \pi_*) O_* = I_* \tag{4}$$

describe the relation between the forcing vector $I_*$ and the resulting response vector $O_*$ in a given experiment. In the frequency-response function vector $O_*$ and in the forcing vector $I_*$ both the mechanical and electrical components, that is, all the degrees of freedom of the electro-mechanical system, are listed. The matrix $D(\omega, \pi_*)$, depending on the frequency and on the actual value of the parameters $\pi_*$, can be obtained by finite-elements procedures or, in the case of simple structures, in exact form by spectral-elements procedures [14]; see the Appendix for further details. Because not all the degrees of freedom of the system are observed, it is useful to consider the following partition of $O_*$ and $I_*$:

$$O_* = \{m_*, n_*\}^T, \quad I_* = \{g_*, h_*\}^T \tag{5}$$

into the measured, $m_*$, and nonmeasured, $n_*$, components of the response and into their dual quantities $g_*$ and $h_*$. Accordingly Eq. (4) is reduced to

$$\tilde{D}(\omega, \pi_*) m_* = g_* - H(\omega, \pi_*) h_* \tag{6}$$

with

$$D(\omega, \pi_*) = \begin{pmatrix} D_{mm} & D_{mn} \\ D_{nm} & D_{nn} \end{pmatrix}, \quad \tilde{D}(\omega, \pi_*) := D_{mm} - D_{mn} D_{nn}^{-1} D_{nm} \tag{7}$$

and $H(\omega, \pi_*) := D_{mn} D_{nn}^{-1}$. Let us remark that even if the functional dependence of the matrix $D$ over the parameter $\pi$ could be linear, the reduction from Eq. (4) to Eq. (6) via Eq. (7) leads to a nonlinear dependence of the matrix $\tilde{D}$ over the parameter $\pi$; this fact, namely the impossibility of measuring all the degrees of freedom of the system, will lead to nonconvex problems.

Hence, the functional to be minimized over the admissible parameters space $\Pi$ to identify the actual values $\pi_*$ may be chosen as follows:

$$\mathfrak{E}(\pi, m_*, g_*, h_*) = \sum_{k=1}^{K} \left| \tilde{D}(\omega_k, \pi) m_*(\omega_k) - g_*(\omega_k) + H(\omega_k, \pi) h_*(\omega_k) \right|^2 \tag{8}$$

The sum over a set of frequencies $\omega_k$ is crucial to include, in the functional, information over a large frequency bandwidth. Clearly, because of Eq. (6), the function (8) vanishes when $\pi = \pi_*$; moreover 𝔈 is continuous with respect to the parameter vector $\pi$ because it involves continuous functions of $\pi$. When $\pi \neq \pi_*$, the balance equation (6) is not satisfied and the function weighs the so-called unbalanced generalized forces.

**Required Properties of $\mathfrak{E}$: Main Difficulties and Proposed Solutions**

The previous considerations allow us to highlight the main obstacles connected to damage-detection or system-identification problems and eventually to prompt efficient shortcuts. The main difficulties are as follows: a) both the responses and the applied forcing inputs, because of practical impediments, are restricted to a low number of sites: this limits the effectiveness of the functional delegate to ascertain the difference between the undamaged and damaged responses; and b) the measured responses represent global knowledge in the sense that combines local contributions from the overall structure; thus is usually difficult to extract local informations on the actual values of the parameters.

From a mathematical viewpoint, this is tantamount to say that the functional $\mathfrak{E}(\cdot; I_*, O_*)$ can manifest a low sensitivity to variations of the parameters, that is,

$$|\mathfrak{E}(\pi; I_*, O_*) - \mathfrak{E}(\pi_*; I_*, O_*)| < \varepsilon \quad \text{for } \pi \in \Pi \tag{9}$$

with $\varepsilon$ a positive number measuring the experimental sensitivity.

To overcome the depicted drawbacks, it is here proposed to couple the main structure (ms), whose mechanical properties have to be detected, with an auxiliary electric circuit (aec). The coupling between the two subsystems (ms and aec) is ensured by distributing an array of piezoelectric transducers along the structure. Therefore, for a suitable choice of the auxiliary electric circuit the electric response to any kind of forcing inputs is influenced by the mechanical constitutive properties, and one can detect structural damages through purely electric inputs and measurements. To this aim a fundamental hypothesis, which is not specific to dealing with electric components, is the perfect knowledge of the electric constitutive parameters.

The proposed augmented system resolves some of the aforementioned difficulties:

1. It allows us to easily measure the frequency-response functions of several sites or, to be more exact, of several degrees of freedom of the electric circuit; indeed, because the piezoelectric patches are supposed to be uniformly distributed along the structure, the associated electric degrees of freedom are in one-to-one correspondence with the structural regions where they are glued.
2. A proper choice of the function $\mathfrak{E}(\pi; I_*, O_*)$ satisfies the conditions (2). The chosen function will weigh the frequency-response amplitudes in several fixed values $\omega_k$ of the frequency.

3. It sensibly reduces the value $\varepsilon$ of the experimental sensitivity, the right-hand term in Eq. (9). Indeed the electric voltages (or currents) can be measured within tolerances much lower than the typical tolerances of structural eigenfrequencies.
4. Moreover, the constitutive parameters of the auxiliary circuit can be tuned to enhance the sensitivity of the chosen function $\mathfrak{E}$, namely the left-hand term in Eq. (9).

This last property of the proposed auxiliary circuit deserves more attention and, as will be shown in the following, is very useful: the set $\Pi$ of the constitutive parameters of interest is the disjoint union of both the mechanical parameters of the main subsystem ($\Pi_{ms}$) and the electric parameters of the auxiliary subsystem ($\Pi_{aec}$). Suppose a further partition of $\Pi$

$$\Pi = \Pi_1 \cup \Pi_2, \quad \Pi_1 \cap \Pi_2 = \emptyset \qquad (10)$$

into the set $\Pi_1$ of parameters to be identified and the set $\Pi_2$ of the parameters that can be used to maximize the sensitivity of the function with respect to the parameters in $\Pi_1$. The actual value $\pi_*$ of all the system parameters in the experiment can be consequently written as $\pi_* = \{\pi_{1*}, \pi_{2*}\}$; $\pi_{1*}$ is the actual value of the parameters to be identified and $\pi_{2*}$ is the actual value of the remaining parameters. Because of the previous hypotheses, the parameters in the set $\Pi_2$ are supposed perfectly known. If at least some of them can be easily controlled (for instance the electric parameters in the circuit), then one can use these to enhance the sensitivity of the function $\mathfrak{E}$. In Fig. 1 we exemplify this enhancement procedure by showing a possible plot and contour plot for the function $\mathfrak{E}$: if a value $\pi_{2*}$ exists for which the restricted function $\mathfrak{E}\left(\cdot, \pi_{2*}; I_*, O_*\right)$ resolves more sharply the actual value $\pi_{1*}$ then $\pi_{2*}$ has to be chosen in the experiment to increase the detection sensitivity. The entire process for the identification of the actual unknown parameters ($\pi_{1*}$ and $\pi_{2*}$) would consist of a sequence of maximization and minimization problems according to the following scheme:

$$\begin{aligned} \pi_{2,k+1} = \pi_2 \quad \text{to achieve} \quad \max_{\pi_2 \in \Pi_2} \mathfrak{E}(\pi_{1k}, \pi_2; I_*, O_*) \\ \pi_{1,k+1} = \pi_1 \quad \text{to achieve} \quad \min_{\pi_1 \in \Pi_1} \mathfrak{E}(\pi_1, \pi_{2,k+1}; I_*, O_*) \end{aligned} \quad (11)$$

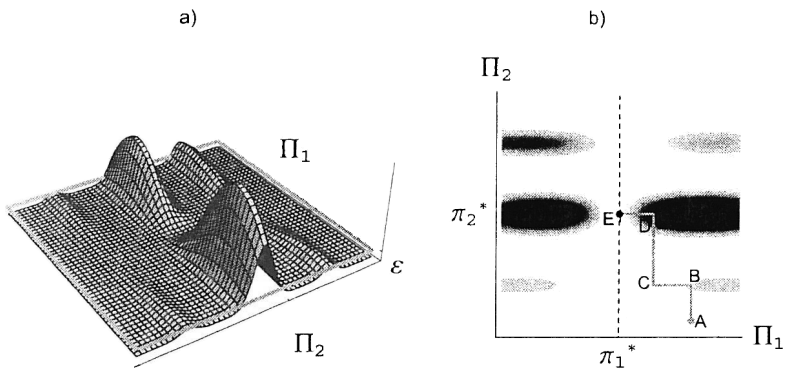

**FIGURE 1.** Min–max process associated with the use of an auxiliary subsystem in damage detection.

The overall procedure is illustrated in Fig. 1 where the different maximization (AB, CD) and minimization (BC, DE) steps are drawn.

The choice $\Pi_1 = \Pi_{ms}$, $\Pi_2 = \Pi_{aec}$ compels the use of the electric auxiliary parameters to maximize the functional sensitivity to the mechanical parameters but, obviously, it is not the only one conceivable. We explicitly remark that the scheme (11) does not converge in every region of the parameters and for every damage-detection function, several locally optimal points can in general exist; however it converges if the region is restricted to be close enough to the global optimal solution. For this reason to avoid problems with multiple optima, the use of a coarse estimation of the actual damage parameters is compelled.

## DESCRIPTION OF THE OVERALL SYSTEM

The proposed method for structural-damage identification relies on the coupling of the main structure with an auxiliary electric network. The energy is transformed from the mechanical to the electric form by means of a set of piezoelectric patches distributed along the structure. These kind of electromechanical devices has been initially proposed (e.g., Ref. 15) to control structural vibrations. In Fig. 2 a sketch of a PEM structure is shown; the piezoelectric patches and the passive electric network connecting them constitute an auxiliary system. More precisely, from the electric viewpoint, the piezoelectric patches behave as capacitors in parallel with current generators (the current being proportional to the time rate of local mechanical strain); from the mechanical viewpoint, the same patches are local stiffeners at the end

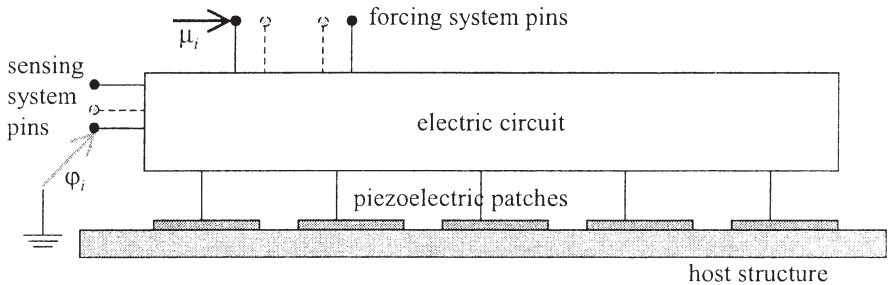

**FIGURE 2.** PEM structure: host structure, piezo patches, electric circuit, and forcing and sensing systems.

points of which two opposite couples proportional to the electric voltage are applied.

The time evolution of these electromechanical systems are described by the functions

$$u \in \mathcal{U}, \quad \varphi_i \in \mathcal{F}, \quad i = 1, 2, \ldots, N$$

meaning respectively the displacement fields and the flux linkages (i.e., time integrals of the electric potentials) of the $N$ nodes in the circuit. Here $\mathcal{U}$ and $\mathcal{F}$ means two suitable functional spaces respectively; to fix ideas one could set $\mathcal{U} = H_2(D \times T, \mathrm{IR}^M)$ and $\mathcal{F} = H_2(T, \mathrm{IR})$, with $D$ being the reference domain of the structure and $T$ the time axis. The evolution equations read as follows:

$$\begin{cases} A(u) + \ddot{u} - g_i(x)\dot{\varphi}_i = F \\ \beta_{ij}\varphi_j + \ddot{\varphi}_i + \delta_{ij}\dot{\varphi}_j + \gamma_{ij}G(\dot{u})|_{x_j} = \mu_i, \quad i = 1, 2, \ldots, N \end{cases} \quad (12)$$

where $A$ is a linear self-adjoint differential operator, $\dot{\varphi}_i$ the electric potential in the $i$th of the $N$ electric nodes, and ($\mathcal{F}$ and $\mu_i$ the applied loads and the applied current acting on the $i$th electric node. If the circuit is realized using only resistances, inductances, and transformers, the matrices $\beta_{ij}$ and $\delta_{ij}$ are guaranteed to be symmetric and positive defined. In Eq. (12) the superposed dot means time differentiation and summation over repeated indices. The linear differential operator $G$, the matrix $\gamma_{ij}$, and the $N$-component vector $g_i(x)$ account for the piezoelectric couplings. Here and in what follows the involved physical quantities have been made dimensionless following the standard practice in designing experimental setups; as reference quantities the beam length, the first natural period of the uncoupled mechanical system, and the electric flux linkage $\bar{\varphi} = \sqrt{M_B/C_N}$, where $M_B$ is the beam mass per unit length and $C_N$ the capacitance of the electric network per unit length, have been chosen (see, for instance, Ref. 16).

Once the number of piezoelectric patches is fixed, one can attempt an optimization procedure on the coefficients $\beta_{ij}$ and $\delta_{ij}$ to maximize the sensitivity of the overall system to the even local changes of the mechanical constitutive parameters (in this case reflected in alterations of the operator $A$). Different circuit topologies are possible to connect the piezoelectric patches among them, the selected one being indicated either by optimal conditions found for $\beta_{ij}$ and $\delta_{ij}$ or simply by simplicity requests. For instance the simplest one is realized by choosing a circuit with only one degree of freedom as done in Fig. 3.

In this case the optimization procedure, to enhance the sensitivity to local changes of mechanical stiffness, is applied to the scalar parameters $\beta$ and $\delta$, meaning respectively the impedance and resistance in the circuit. It is useful to remark that this circuital scheme is a simple generalization of the 1-degree-of-freedom (DOF) shunt circuit by Hagood and von Flotow [17], a standard technique for vibration control.

110 F. dELL'ISOLA ET AL.

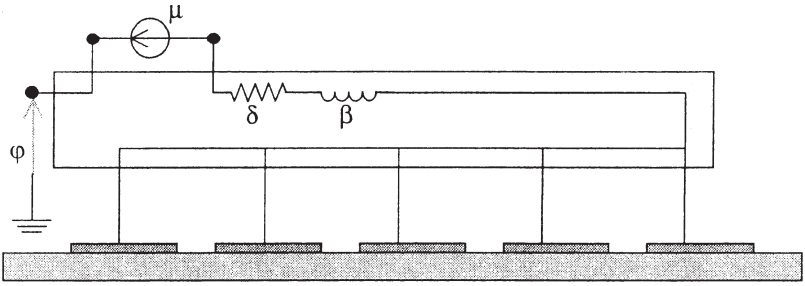

**FIGURE 3.** Simplest auxiliary electric circuit: 1-degree-of-freedom shunt circuit.

In the Conclusions, the auxiliary electric circuit is chosen instead as the analog circuit of the structure to be detected, namely an Euler–Bernoulli beam; thus the network is governed, in the homogenized limit, by a fourth-order space derivative (see, e.g., Ref. 18 for details).

The discussed procedure of damage identification is here applied to the relevant case of a simply supported Euler–Bernoulli beam. This choice represents the simplest experimental setup that can be conceived to prove the feasibility of the proposed sensitivity-enhancement method.

### Transmission Line as Auxiliary System

As the auxiliary electric circuit connecting the piezoelectric patches distributed along the beam the fourth-order transmission line synthesized in Ref. 18 has been chosen. Two moduli of this connection are shown in Fig. 4; in this case an inductance and a transformer are needed in each module; however, through the use of multiple channel date-acquisition boards, these circuits can also be simulated by computer.

Once an homogenization procedure has been carried out, the governing equations for the proposed augmented electromechanical system, in case of

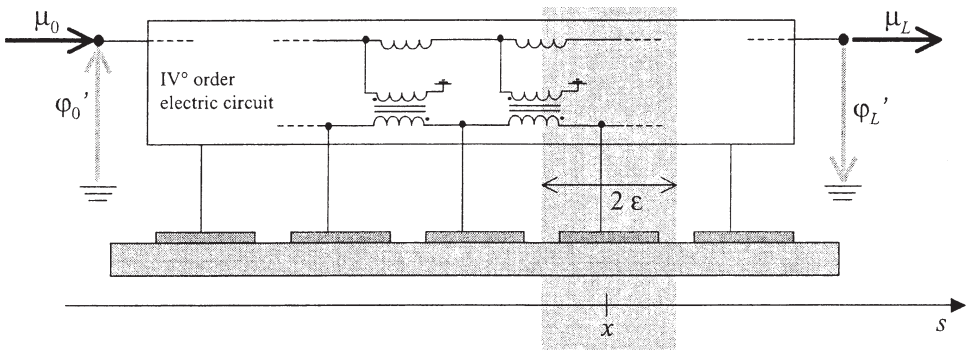

**FIGURE 4.** Euler beam coupled to its electric analog circuit.

nonvanishing forces only at the boundary, can be written as

$$\begin{cases} (\alpha\, u'')'' + \ddot{u} - \gamma \dot{\varphi}'' = 0 & (13a) \\ \beta \varphi^{IV} + \ddot{\varphi} + \delta \dot{\varphi} + \gamma \dot{u}'' = 0 & (13b) \end{cases}$$

for each regular interval in [0, 1]. Here, and in what follows, a prime means the space derivative and all the quantities are dimensionless. We explicitly remark that the number of piezoelectric patches chosen should be large enough to assure that the continuum model (13) is applicable; a rule of thumb is to distribute at least four patches in the smallest wavelength considered. Of course, if one deals with a limited number of piezoelectric patches, the discrete model would be more reliable, but the continuum model (13) has the advantage of immediately showing the main features of this electro-mechanical system. For instance, the comparison of Eq. (13a) with Eq. (13b) shows that PEM systems are based on internal resonance phenomena; to see this, simply choose $\alpha$ to be constant along the beam and $\beta = \alpha$.

Recalling that $\dot{\varphi}$ physically represents the electric potential, note that the constitutive relation for the dimensionless bending moment in this case reads

$$M = \alpha u'' - \gamma \dot{\varphi} \qquad (14)$$

where $\alpha$ is the bending stiffness and $\gamma$ the coefficient resulting from the piezoelectric coupling. In a similar way, also the *electric bending moment* that is, the electric current expending power on $\dot{\varphi}''$, has two different contributions, namely

$$\mu = \beta \varphi'' + \gamma \dot{u} \qquad (15)$$

The system is subjected to the following boundary conditions:

$$\begin{aligned} u(0,t) = u(1,t) = 0, &\quad M(0,t) = M(1,t) = 0 \\ \varphi(0,t) = \varphi(1,t) = 0, &\quad \mu(0,t) = \mu_0(t); \quad \mu(1,t) = \mu_1(t) \end{aligned} \qquad (16)$$

These correspond to a simply supported beam and to an electrically grounded transmission line that is subjected to a nonvanishing value of the electric bending moment at its edges, $\mu_0$ and $\mu_1$ (Fig. 4); the electric signal used for testing are ideal Dirac deltas in all the following simulations. Because of Eqs. (14) and (15), the essential boundary conditions in Eq. (16) are expressed in terms of kinematical fields:

$$u''(0,t) = u''(1,t) = 0, \qquad \varphi''(0,t) = \mu_0(t)/\beta, \quad \varphi''(1,t) = \mu_1(t)/\beta \quad (17)$$

It has been shown [11] that the proposed circuital scheme guarantees a multi-modal coupling between the beam and the electric system for a proper value of the line inductance; in other words, the tuning of the line inductance allows for the simultaneous internal resonance of all the structural modes with all the electric modes. This circumstance generalizes the technique of Hagood to the case of multimodal control and leads to the most efficient coupling between the mechanical and electric subsystems. Moreover, in this condition the system reveals to be very effective to electrically sense the structural damage.

## Detection and Localization of Structural Stiffness Reduction

With reference to the described system (Fig. 4), we identify the damaged profile of the dimensionless bending stiffness $\alpha$ by means of purely electric measurements (the flux linkages $\varphi'_0, \varphi'_1$) and purely electric forcing inputs (the Dirac deltas $\mu_0$ and $\mu_1$) in the auxiliary network. Indeed, only the following quantities

$$m_* = \{\varphi'_0, \varphi'_1\}^\top, \qquad g_* = \{\mu_0, \mu_1\}^\top, \qquad h_* = 0 \qquad (18)$$

are measured and contribute to the functional $\mathfrak{E}$ to be minimized, namely the functional chosen in Eq. (8). For the dimensionless bending stiffness $\alpha(s)$ along the beam span, we seek for solutions in the form

$$\alpha(s) = \begin{cases} \alpha_0, & \text{for } s \notin (x-\epsilon, x+\epsilon) \\ \alpha_0 d, & \text{for } s \in (x-\epsilon, x+\epsilon) \end{cases} \qquad (19)$$

This hypothesis reduces the space of parameters $\Pi$ to a finite dimensional space and hints to a purely locally ($\epsilon \ll 1$) damaged state of the beam; indeed $d \in [0, 1]$ is the percentage loss with respect to the undamaged state, $x \in (\epsilon, 1-\epsilon)$ the abscissa where the damaged zone is centered, and $2\epsilon$ the range of the damaged zone that is assumed to be known; refer to the gray shaded zone in Fig. 4. Thus, setting

$$\Pi_1 = \{(d, x)/d \in (0, 1), \ x \in (\epsilon, 1-\epsilon)\}, \qquad \Pi_2 = \{\beta/\beta > 0\} \qquad (20)$$

means to seek for the correct value $\pi_{1*} = \{d_*, x_*\}$ of the unknown constitutive mechanical parameters, using the electric parameter $\beta$, constant with respect to the beam abscissa $s$, to increase the functional sensitivity. Because of the simplicity of the geometry, the system is divided into three parts, each with constant parameters, and the spectral element method [14] is used to compute the exact form of the matrix $\tilde{D}(w_k, \pi)$ for all the needed iterates of the maximization–minimization process.

The dimensionless quantities used in the numerical simulations are $\alpha_0 = 1, \gamma = 1/20, \epsilon = 1/20, d_* = 0.5, x_* = 0.8$, and $\delta = 0$. All of them correspond to technically feasible conditions and materials.

To understand how the electric $\Pi_2$-parameter $\beta$ can affect the sensitivity of the function with respect to the $\Pi_1$ parameters, $d$ and $x$, the locus of all the points satisfying $\mathfrak{E}(d, x, \beta) = \varepsilon$ is drawn in Fig. 5, with the experimental sensitivity fixed to the value $\varepsilon = 1$ and $\beta$ ranges from 0.8 to 1.2. Note that the tightened regions in the plot correspond to values of $\beta \simeq 1 = \alpha_0$; this condition, in case of an undamaged structure, would lead to a complete modal coupling of all the mechanical and electrical modes, because all the mechanical and electrical eigenfrequencies of Eq. (13) would match, as previously discussed. In Fig. 5 the points internal to the surface represent the parameter values where the functional $\mathfrak{E}(d, x, \beta)$ is less than $\varepsilon$ or in other words the parameter values that are solution of the identification problem within a tolerance of $\varepsilon$. Thus, the values over the axis $\Pi_2$ where the surface is shrunk correspond to optimal values of the parameter $\beta$ because less variance is allowed for the solutions in the $\Pi_1$ parameters. Figure 5 in itself represents a comparison between the standard and the proposed enhanced evaluation procedure, the advantage of this last relying in the optimization of the damage-detection functional sensitivity; this allows us to get a sharper resolution and as a consequences to obtain a damage characterization within smaller confidence ranges.

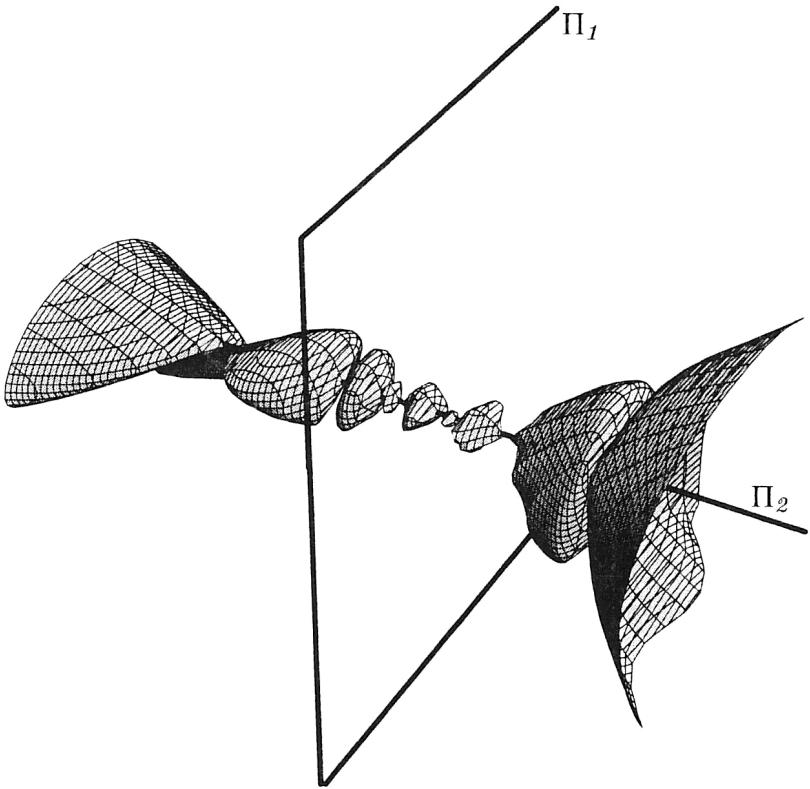

**FIGURE 5.** Points in $\Pi$ satisfying the condition $\mathfrak{E}(d, x, \beta) = \varepsilon$. The parameter $\beta$ ranges from 0.8 to 1.2

In Figs. 6 and 7 for a value of the parameter $\beta = \bar{\beta} = 1$, which results are close to the optimal one, the contour plots of the function $\mathfrak{E}(d, x, \bar{\beta})$ over the parameters space $\Pi_1 = [0.1, 1] \times [0, 1]$ are drawn for two different load conditions. There are several local minima in both cases but only one global minimum. However, in the two different load conditions, the directions of maximum gradient near the global optimum are independent; this means that the two loads give independent information about the damage and can usefully be used in conjunction in the identification process. In both the load cases, disregarding any knowledge about these contour plots, the Nelder–Mead [19] simplex algorithm is successfully used from a set of four random points in $\Pi_1$ to find the global minimum for the given value $\beta = \bar{\beta}$ of the electric parameter. On the other hand, because of the nonconvexity of the functional, the whole min–max process, described in Eq. (11) and leading to the simultaneous identification of the damage parameters and optimal tuning of the electric parameter $\beta$, always converges only if it is started in a convex neighborhood of the point $(d = d^*, x = x^*)$; thus, the min–max procedure is efficient if at least an a priori estimate about the damage is given.

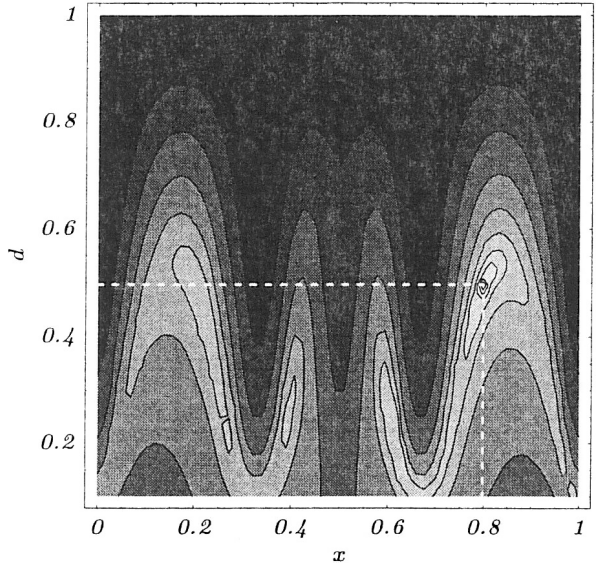

**FIGURE 6.** Level plot of log $(\mathfrak{E}(d, x, \bar{\beta}))$ on the space $\Pi_1$ for the load condition $\mu_0(\omega) = 0$, $\mu_1(\omega) = 1$. The point $d = 0.5$, $x = 0.8$ is the global minimum.

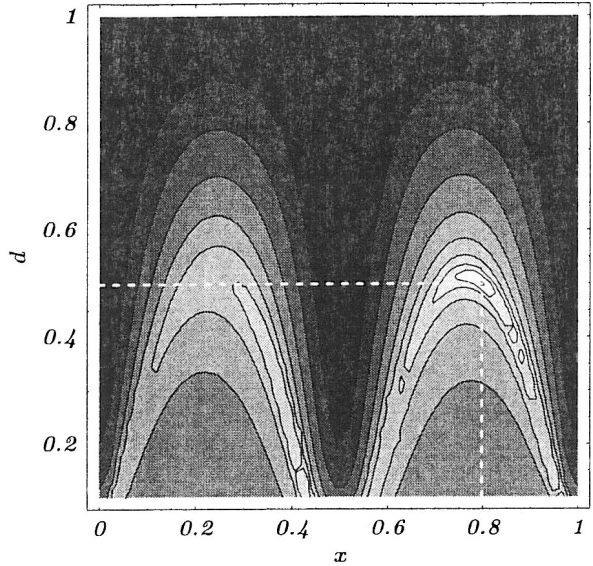

**FIGURE 7.** Level plot of log $(\mathfrak{E}(d, x, \bar{\beta}))$ on the space $\Pi_1$ for the load condition $\mu_0(\omega) = \mu_1(\omega) = 1$.

The chosen damage level, $d_* = 0.5$, and its position, $x_* = 0.8$, lead to variations $\Delta \varpi$ of the first three natural frequencies ranging from 2% to 4%, where the experimental sensitivity $\varepsilon_\varpi$ for natural frequencies measurements is about 1%. On the other side the percentage variations $\Delta \varphi$ of the measured electric flux linkages occurring in the used function can range from 0 to 100% depending on the parameter $\beta$ whereas the associated experimental sensitivity $\varepsilon_\varphi$ is at least 0.1%. Therefore, the comparison of the ratios

$$\frac{\Delta \varpi}{\varepsilon_\varpi} \approx 2 \to 4, \qquad \frac{\Delta \varphi}{\varepsilon_\varphi} \approx 0 \xrightarrow{\beta} 10^3 \qquad (21)$$

meaning the experimental confidences, shows an evident advantage of the proposed technique with respect to the standard method of damage detection, where the variation of frequencies are measured. In particular when the electric subsystem is tuned with $\beta = 1$ the ratio $\Delta \varphi / \varepsilon_\varphi$ is about $10^2$, which represents a very satisfactory result.

## CONCLUSIONS

It is shown how purely electric measurements of voltages in an auxiliary electric circuit allow for the detection of mechanical local damages of a structure. The coupled electro-mechanical system, with respect to the techniques proposed in the literature, usually based on measurements of mechanical eigenfrequencies, turns out to be:

- More flexible, because the selected circuital topology can be easily adapted to satisfy different optimal conditions;
- More sensitive to local changes of mechanical parameters, because it is based on purely electric measurements and the auxiliary circuit is optimized to enhance the sensitivity; and

- More easily tunable, because the electric parameters can be tuned in an adaptive way.

The measurement of the frequency-response function of only two electric degrees of freedom turned out to be sufficient to identify the level and localization of a concentrated damage. Further research efforts should be devoted in determining a procedure of coarse localization of damage to avoid the difficulties caused by multiple optima. The development of a sensitivity-enhancement procedure based on the possibility of an independent tuning of all the inductances appearing in the auxiliary network could be another area of interest; in this way we expect to obtain additional information on the position of the damaged zone while performing the sensitivity tuning.

## APPENDIX: SPECTRAL ELEMENT ASSOCIATED WITH EQ. (13)

The spectral-element (SE) method, as formulated for instance in Ref. 5, is applied to the Eqs. (13); this method is applied when the coefficients of a PDE are piecewise constants and is based on the partition of the domain of interest into suitable subdomains. Each of these becomes a spectral element and the considered PDE is solved, representing its solution in terms of its eigenfunctions in each element. According to the choice (19) of the dimensionless bending stiffness $\alpha(s)$, we divide the support in three parts with constants parameters, namely the intervals $[0, x - \epsilon), (x - \epsilon, x + \epsilon)$ and $(x + \epsilon, 1]$. In each of them, the Fourier transform of Eqs. (13) reads as follows:

$$\begin{cases} \alpha_h u^{IV} - \omega^2 u - i\omega\gamma\varphi'' = 0, \\ \beta\varphi^{IV} - \omega^2 \varphi + i\omega\delta\varphi + i\omega\gamma u'' = 0, \end{cases} \quad h = 1, 2, 3 \quad \text{(A1)}$$

where $\alpha_h$ is the constant value of the dimensionless bending stiffness in the $h$th interval. The SE method seeks for solutions $(u(s, \omega)$ and $\varphi(s, \omega))$ of Eqs. (A1) in the form:

$$u(s, \omega) = \sum_{j=1}^{J} u_{hj} \exp[k_{hj}(\omega)s], \quad \varphi(s, \omega) = \sum_{j=1}^{J} \varphi_{hj} \exp[k_{hj}(\omega)s], \quad \text{(A2)}$$

where the $k_{hj}(\omega)$ functions are the $J$ solution ($J = 8$ in this case) to the dispersion relation in the $h$th interval:

$$\det \begin{pmatrix} \alpha_h k^4 - \omega^2 & -i\omega\gamma k^2 \\ i\omega\gamma k^2 & \beta k^4 - \omega^2 + i\omega\delta \end{pmatrix}$$
$$= \beta\alpha_h k^8 + [i\delta\omega\alpha_h - (\beta + \alpha_h + \gamma^2)\omega^2]k^4 + \omega^4 - i\delta\omega^3. \quad \text{(A3)}$$

Because the dispersion relations are quadratic in $k^4$, they are easily solvable, to get

$$\{k_{hj}(\omega)\} = \{1, -1, i, -i\}$$
$$\times \frac{(\beta + \alpha_h + \gamma^2)\omega^2 - i\delta\omega\alpha_h \pm \sqrt{[(\beta + \alpha_h + \gamma^2)\omega^2 - i\delta\omega\alpha_h]^2 - 4\beta\alpha_h(\omega^4 - i\delta\omega^3)}}{2\beta\alpha_h}$$

(A4)

Accordingly the coefficients $u_{hj}$ and $\varphi_{hj}$ are then chosen to satisfy the boundary conditions of Eq. (16), and the continuity conditions and the jump conditions at the boundaries of the three intervals. The resulting linear system of equations reads as Eq. (4) once the coefficients $u_{hj}$ and $\varphi_{hj}$ are grouped into the output vector $O$ while the boundary applied forces and currents are grouped into the input vector $I$.

Note that, when using the SE method, the dimension of the resulting system (4) is relatively small; for instance in our case (two second gradient equations over three intervals with constant parameters) we have $2 \times 2 \times (3+1) = 16$ degrees of freedoms. Using a FE approach to the same problem would lead to a much bigger problem to get a similar precision. However in SE approach the functional dependence of the dynamic stiffness matrix $D_{SE}(\omega, \pi)$ with respect to the frequency $\omega$ is transcendent—through the exponential functions in Eq. (A2) and the $k_{hj}(\omega)$ functions in Eq. (A4)—whereas in the FE method is polynomial:

$$D_{FE}(\omega, \pi) = K(\pi) - \omega^2 M(\pi) + I\omega C(\pi) \quad (A5)$$

where $K$, $M$, and $C$ are the stiffness, mass, and damping matrices respectively.

In the process of minimizing the damage-detection functional, we need to evaluate $\mathfrak{E}(\pi)$ in many points of the parameters space $\Pi$; for each one of these evaluations, the dynamic stiffness matrix $D_{SE}(\omega, \pi)$ or $D_{FE}(\omega, \pi)$ must be assembled. Because, generally, the assembly of $D_{FE}(\omega, \pi)$ is faster than the assembly $D_{FE}(\omega, \pi)$, the SE method should be preferred. However, there are procedures where the damage-detection functional only accounts for the resonant frequencies [i.e., the singular frequencies of $D(\omega, \pi)^{-1}$]; in these cases the relatively ease and velocity of the SE assembly are lost because the resonant frequencies are found as the roots of a transcendent function, namely the determinant of $D_{SE}(\omega, \pi)$, whereas in the FE case (26) they are found by standard eigenvalue problems.